\documentclass[showpacs,showkeys,prb,twocolumn]{revtex4}
\usepackage{amssymb}
\usepackage{amsfonts}
\usepackage{amsmath}
\usepackage{epsfig}
\usepackage{graphicx}

\providecommand{\U}[1]{\protect\rule{.1in}{.1in}}

\begin{document}

\title{Collective excitations of an imbalanced fermion gas in a 1D optical
lattice}
\author{R. Mendoza}
\affiliation{Posgrado en Ciencias F\'{\i}sicas, UNAM; Instituto de F\'{\i}sica, UNAM}
\author{Mauricio Fortes}
\affiliation{Instituto de F\'{\i}sica, UNAM, Apdo. Postal 20-364, 01000 M\'exico D.F.,
M\'exico}
\author{M. A. Sol\'{\i}s}
\affiliation{Instituto de F\'{\i}sica, UNAM, Apdo. Postal 20-364, 01000 M\'exico D.F.,
M\'exico}


\begin{abstract}
The collective excitations that minimize the Helmholtz free energy of a
population-imbalanced mixture of a $^{6}$Li gas loaded in a quasi one-dimensional
optical lattice are obtained. These excitations reveal a rotonic
branch after solving the Bethe-Salpeter equation under a generalized random
phase approximation based on a single-band Hubbard Hamiltonian. The phase
diagram describing stability regions of Fulde-Farrell-Larkin-Ovchinnikov and
Sarma phases is also analyzed.

\end{abstract}

\maketitle

\section{Introduction}

Optical lattices are tailored made to study strongly-correlated Fermi
systems, the stability of different phases and the effects of dimensionality
in population-imbalan ced mixtures of different species of ultra cold gases
under attractive interactions \cite{Esslinger, Ketterle, Ketterle2,  Liao}. In the
latter case, the Fermi surfaces of each species are no longer aligned and
Cooper pairs have non-zero total momenta $2\bold{q}$. Such phases were
first studied by Fulde and Ferrell (FF) \cite{FF}, who used an order
parameter that varies as a single plane wave, and by Larkin and Ovchinnikov
(LO) \cite{LO}, who suggested that the order parameter is a superposition of
two plane waves.

The mean-field treatment of the FFLO phase in a variety of systems, such as
atomic Fermi gases with population imbalance loaded in optical lattices \cite%
{Kop1,Kop,FG,Machida,Capone,KMF,KMF2}, shows that the FFLO phase competes
with a number of other phases, such as the Sarma ($\boldsymbol{q}=0$) states 
\cite{S, Wilczek}, but in some regions of momentum space the FFLO phase is more
stable as it provides the minimum of the mean-field expression of the
Helmholtz free energy. In addition, recent calculations on the FFLO phase of
the same system in two- and three-dimensional\cite{KMF,KMF2} optical lattices suggest that
the region of stability of this phase as a function of polarization
increases when the dimensionality of the periodic lattice is lowered.

In this work, we use a Bethe-Salpeter approach to obtain the collective excitations of the two-particle propagator of a polarized mixture
of two hyperfine states $|\!\!\uparrow >$ and $|\!\!\downarrow >$ of a $^{6}$ Li atomic Fermi gas with attractive interactions loaded in a quasi one-dimensional optical lattice described by a single-band Hubbard Hamiltonian. 

\section{Hubbard model}

The Hamiltonian of a two-component Fermi gas under an attractive contact
interaction in a periodic lattice with constant $a$ is \cite{KMF}%
\begin{eqnarray}
H &=&- J_{x}\sum_{\left\langle i,j\right\rangle _{x},\sigma }\hat{c}%
_{i,\sigma }^{\dag }\hat{c}_{j,\sigma }-J_{y}\sum_{\left\langle
i,j\right\rangle _{y},\sigma }\hat{c}_{i,\sigma }^{\dag }\hat{c}_{j,\sigma} \nonumber \\
&& - J_{z}\sum_{\left\langle i,j\right\rangle _{z},\sigma }\hat{c}_{i,\sigma
}^{\dag }\hat{c}_{j,\sigma }
 - \sum\limits_{i}\left( \mu _{\uparrow }^{\dag }%
\hat{c}_{i,\uparrow }^{\dag }\hat{c}_{i,\uparrow }+\mu _{\downarrow }\hat{c}%
_{i,\downarrow }^{\dag }\hat{c}_{i,\downarrow }\right)
  \nonumber \\
&& + U\sum\limits_{i}\hat{c}_{i,\uparrow }^{\dag }\hat{c}_{i,\downarrow
}^{\dag }\hat{c}_{i,\downarrow }\hat{c}_{i,\uparrow },  \label{Hub}
\end{eqnarray}%
where $J_{\nu }$ is the tunneling strength of the atoms between
nearest-neighbor sites in the $\nu $-direction; $U$ is the on-site
attractive interaction strength; $\mu _{\uparrow ,\downarrow }$ is the
chemical potential of species $|\!\!\uparrow >$, $|\!\!\downarrow >$, and the
Fermi operator $\hat{c}_{i,\sigma }^{\dag }$ ($\hat{c}_{i,\sigma }$) creates
(destroys) an atom on site $i$. We assume a system with a total number of
atoms $M=M_{\uparrow }+M_{\downarrow }$ distributed along $N$ sites of the
optical-lattice potential. For a quasi one-dimensional (1D) system the
tunneling strengths satisfy $J_{x}\gg J_{y}=J_{z}$ and the usual tight-binding
lattice dispersion energies are $\xi _{\uparrow ,\downarrow }(\boldsymbol{k}%
)=2\sum_{\nu }J_{\nu }\left( 1-\cos k_{\nu }a\right) -\mu _{\uparrow
,\downarrow}$.

The order parameter $\Delta _{i}=U\left\langle \hat{c}_{i,\downarrow }\hat{c}%
_{i,\uparrow }\right\rangle $ of the FFLO states is assumed to vary as a single plane
wave, $\Delta _{i}=\Delta \exp \left( 2\imath \bold{q}\cdot 
\bold{r}_{i}\right) $, where 2$\bold{q}$ is the pair
center-of-mass momentum,  $\bold{r}_{i}$ the coordinate of site $i$, and $\Delta$ is the usual BCS gap.

\section{Phase diagrams}

Within the mean field approximation and using a Bogoliubov transformation to
diagonalize the Hamiltonian, the grand canonical partition function $Z$ can
be obtained in terms of both, electronlike and holelike dispersion $\omega
_{\pm }=E_{\boldsymbol{q}}(\boldsymbol{k})\pm \eta _{\boldsymbol{q}}(%
\boldsymbol{k}),$ where $\eta _{\boldsymbol{q}}(\boldsymbol{k})=\frac{1}{2}%
\left[ \xi _{\uparrow }(\boldsymbol{k+q})-\xi _{\downarrow }(\boldsymbol{q-k}%
)\right] $ and $E_{\boldsymbol{q}}(\boldsymbol{k})=\sqrt{\chi _{\boldsymbol{q%
}}^{2}(\boldsymbol{k})+\Delta ^{2}}.$ The thermodynamic potential $\Omega =-%
\frac{1}{\beta }\ln Z$ is 
\begin{eqnarray}
\Omega  &=&\frac{1}{N}\sum_{\boldsymbol{k}}\left[ \chi _{\boldsymbol{q}}(%
\boldsymbol{k})+\omega _{-}(\boldsymbol{k},\boldsymbol{q})+\frac{\Delta ^{2}%
}{U}\right]   \nonumber \\
&&\hspace{-1.5cm}-\frac{1}{\beta }\sum_{\boldsymbol{k}}\left[ \ln \left(
1+e^{-\beta \omega _{+}(\boldsymbol{k},\boldsymbol{q})})+\ln (1+e^{\beta
\omega _{-}(\boldsymbol{k},\boldsymbol{q})}\right) \right] ,
\label{thermpot}
\end{eqnarray}%
where $\chi _{\boldsymbol{q}}(\boldsymbol{k})=\frac{1}{2}\left[ \xi
_{\uparrow }(\boldsymbol{k+q})+\xi _{\downarrow }(\boldsymbol{q-k})\right] $
and $\beta =1/k_{B}T.$ The Helmholtz free energy $F(\Delta ,\boldsymbol{q}%
,f_{\uparrow },f_{\downarrow },T)=\Omega +\mu _{\uparrow }f_{\uparrow }+\mu
_{\downarrow }f_{\downarrow },$ where $f_{\uparrow ,\downarrow }\equiv
M_{\uparrow ,\downarrow }/N$ is the spin-up (spin-down) filling fraction,  can now be minimized with respect to $\Delta ,$ 
$\boldsymbol{q},$ $\mu _{\uparrow }$ and $\mu _{\downarrow }.$ This leads to
a set of four equations that determine the stable phases of this system as a
function of temperature and polarization $P\equiv \frac{f_{\uparrow
}-f_{\downarrow }}{f_{\uparrow }+f_{\downarrow }}.$%
\begin{align}
& f_{\uparrow }=\frac{1}{N}\sum_{\boldsymbol{k}}\left[ u_{\boldsymbol{q}%
}^{2}(\boldsymbol{k})f(\omega _{+}(\boldsymbol{k},\boldsymbol{q}))+v_{%
\boldsymbol{q}}^{2}(\boldsymbol{k})f(-\omega _{-}(\boldsymbol{k},\boldsymbol{%
q}))\right] ,  \nonumber \\
& f_{\downarrow }=\frac{1}{N}\sum_{\boldsymbol{k}}\left[ u_{\boldsymbol{q}%
}^{2}(\boldsymbol{k})f(\omega _{-}(\boldsymbol{k},\boldsymbol{q}))+v_{%
\boldsymbol{q}}^{2}(\boldsymbol{k})f(-\omega _{+}(\boldsymbol{k},\boldsymbol{%
q}))\right] ,  \nonumber \\
& 1=\frac{U}{N}\sum_{\boldsymbol{k}}\frac{1-f(\omega _{-}(\boldsymbol{k},%
\boldsymbol{q}))-f(\omega _{+}(\boldsymbol{k},\boldsymbol{q}))}{2E_{%
\boldsymbol{q}}(\boldsymbol{k})}  \nonumber \\
& 0=\frac{1}{N}\sum_{\boldsymbol{k}}\left\{ \frac{\partial \eta _{%
\boldsymbol{q}}(\boldsymbol{k})}{\partial q_{x}}\left[ f(\omega _{+}(%
\boldsymbol{k},\boldsymbol{q}))-f(\omega _{-}(\boldsymbol{k},\boldsymbol{q}))%
\right] +\frac{\partial \chi _{\boldsymbol{q}}(\boldsymbol{k})}{\partial
q_{x}}\right.   \nonumber \\
& \!\times \!\left. \left[ 1-\frac{\chi _{\boldsymbol{q}}(\boldsymbol{k})}{%
E_{\boldsymbol{q}}(\boldsymbol{k})}\left[ 1-f(\omega _{+}(\boldsymbol{k},%
\boldsymbol{q}))-f(\omega _{-}(\boldsymbol{k},\boldsymbol{q}))\right] \right]
\right\} ,  \label{foureqs}
\end{align}%
where $u_{\boldsymbol{q}}(\boldsymbol{k})=\sqrt{\frac{1}{2}\left[ 1+\frac{%
\chi _{\boldsymbol{q}}(\boldsymbol{k})}{E_{\boldsymbol{q}}(\boldsymbol{k})}%
\right] }$, $v_{\boldsymbol{q}}(\boldsymbol{k})=\sqrt{\frac{1}{2}\left[ 1-%
\frac{\chi _{\boldsymbol{q}}(\boldsymbol{k})}{E_{\boldsymbol{q}}(\boldsymbol{%
k})}\right] }$ and $f(x)$ is the Fermi distribution function.

\begin{figure}[tbh]
\begin{center}
\includegraphics[  width=0.75\linewidth,
keepaspectratio]{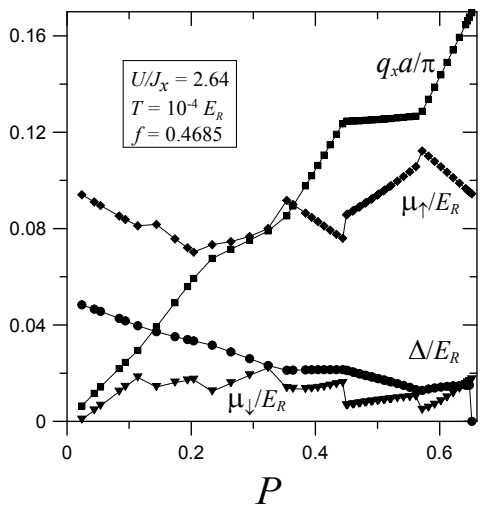}
\end{center}
\par
\caption{Gap (dots), chemical potentials and pair momentum that minimize the Helmholtz free energy.}
\label{fig:F2}
\end{figure}

In Fig. \ref{fig:F2} we show the gap $\Delta $, pair momentum $q_{x}~$and chemical potentials for each species $\mu _{\uparrow },$ $\mu _{\downarrow }$
as a function of polarization that minimize the Helmholtz free energy for a
total filling factor $f=f_{\uparrow }+f_{\downarrow }=0.4685$; hopping
strengths $J_{x}=0.078\ E_{R},$ $J_{y}=J_{z}=10^{-4}E_{R},$ and $U/Jx=2.64$,
where $E_{R}=\hbar ^{2}(2\pi /\lambda )^{2}/2m$ is the recoil energy and $%
\lambda =1030$ nm is the lattice wavelength. When $P\neq 0,$ $q_{x}\neq 0$
and $\Delta \neq 0$ the system is in the FFLO phase which becomes the BCS
state when both $P\rightarrow 0$ and $q_{x}\rightarrow 0$; when $P\neq 0,$
but $q_{x}=0$ the system is in the Sarma \cite{S} or breached-pair phase 
\cite{Wilczek} and the transition to the normal state occurs when the gap $%
\Delta $ vanishes.

\begin{figure}[tbh]
\begin{center}
\includegraphics[  width=0.85\linewidth,
  keepaspectratio]{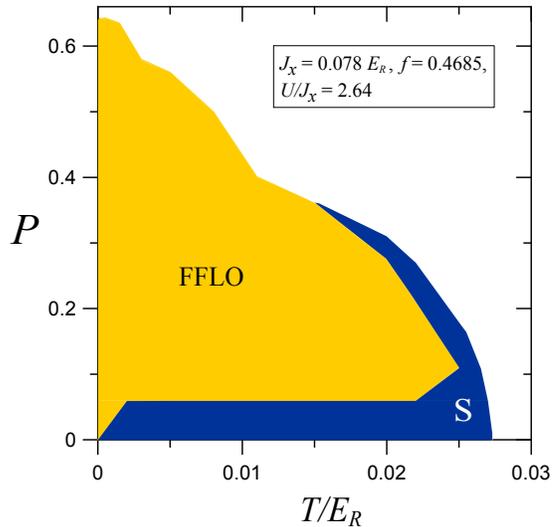}
\end{center}
\par
\caption{(Color online) Phase diagram of an imbalanced Fermi gas in a quasi 1D optical lattice.}
\label{fig:phdiag}
\end{figure}

The phase diagram of the quasi 1D system is shown in Fig \ref{fig:phdiag}. It is interesting to note that the FFLO phase is dominant over quite a large region of the phase-diagram and is stable at higher values for the polarization (up to $P\simeq {0.64}$) compared to our previous work in 2D\cite{KMF} and 3D\cite{KMF2} systems with the same composition and dynamical parameters. In
addition, the mixed phase or phase separation regime in which an unpolarized
BCS core and a polarized normal fluid (in momentum space) coexist at very low temperatures and moderate polarizations \cite{Cald} is no longer present in this system in contrast to the 3D system (and to a lesser extent in the 2D optical lattice).

\section{Collective states}

The spectrum of the collective modes can be obtained from the poles of the
two-particle Green's function $K(1,2;3,4),$ where we use the compact
notation $1=\{\sigma _{1},\boldsymbol{r}_{1},t_{1}\},$ $2=\{\sigma _{2},%
\boldsymbol{r}_{2},t_{2}\},...$ with $\sigma _{i}$ denoting the spin
variables, $\boldsymbol{r}_{i}$ the vector for lattice site $i$, and $t_{i}$%
, the time variable. $K$ satisfies the following Dyson equation:%
\begin{equation}
K=K_{0}+K_{0}IK,  \label{BSdyson}
\end{equation}%
where $K_{0}(1,2;3,4)$ is the two-particle free propagator which is defined
by a pair of fully dressed single-particle Green%
\'{}%
s function,%
\[
K_{0}(1,2;3,4)=G(1;3)G(4;2),
\]%
and the interaction kernel $I$ is given by functional derivatives of the mass operator. 

Using the generalized random phase approximation, we replace the single-particle
excitations with those obtained by diagonalizing the
Hartree-Fock (HF) Hamiltonian while the collective modes are obtained by
solving the Bethe-Salpeter (BS) equation in which the single-particle
Green's functions are calculated in the HF approximation, and the BS kernel
is obtained by summing ladder and bubble diagrams\cite{ZK}. The resulting
equation for the BS amplitudes $\hat{\Psi}_{q}(k,\boldsymbol{Q})$ is 
\begin{equation}
\hat{\Psi}_{q}(k,Q)=-U\hat{D}\sum_{p}\hat{\Psi}_{q}(p,Q)+U\hat{M}\sum_{p}%
\hat{\Psi}_{q}(p,Q),  \label{BSE}
\end{equation}%
where $\hat{\Psi}_{q}(k,\boldsymbol{Q})$ is a
vector with four components and the $4\times 4$ matrices $U\hat{D}$ and $U%
\hat{M}$ represent the contribution resulting from the direct and exchange interactions, respectively. The
dispersion $\omega (\boldsymbol{Q})$ for the collective excitations is
obtained from the solutions of the $4\times 4$ secular determinant defined
by (\ref{BSE}).
\begin{figure}[tbh]
\begin{center}
\includegraphics[  width=0.85\linewidth,
  keepaspectratio]{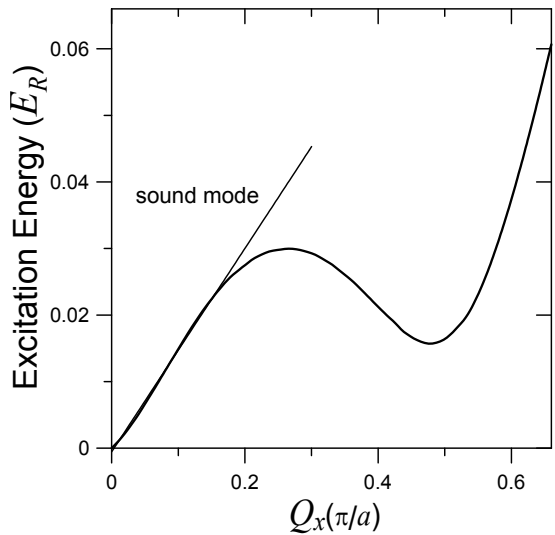}
\end{center}
\par
\caption{Excitation energy for collective modes of a polarized $^6$Li gas in a quasi 1D optical lattice with $\protect\lambda =1030$ nm and total filling factor $f=0.4685.$ The Hubbard parameters are $J_{x}=0.078\ E_{R}$, $J_{y}=J_{z}=10^{-4} \ E_{R}$ and the attractive on-site interaction is $U/J=2.64.$}
\label{fig:rot1D}
\end{figure}
In Fig. \ref{fig:rot1D} we show the collective excitations of the 1D
system. For small $\boldsymbol{Q}$ the Goldstone mode is clearly present
with a sound velocity of $v_{s}=4.92$ mm/s.  For larger $\boldsymbol{Q}$, a
rotonlike minimum appears with a gap $\Delta _{r}=0.0157E_{R}$ and a critical flow velocity $v_{f}=1.05$ mm/s.

\section{Conclusions}
We have shown that superfluid phases of the FFLO and Sarma types are present in ultra cold Fermi gases loaded in quasi one-dimensional optical lattices. The region of stability of the FFLO states in the phase diagram is larger and supports higher population imbalances than identical systems in 2D and 3D. The energy dispersion of collective excitations have the usual Goldstone-mode behavior with a sound velocity of 4.92 mm/s. In addition, for higher momenta a rotonic branch is also present.

\begin{acknowledgements}
We acknowledge the partial support from UNAM-DGAPA grants IN105011,  IN-111613 and CONACyT grant 104917.
\end{acknowledgements}




\end{document}